\documentclass{article}
\usepackage{spconf}
\usepackage{microtype}
\usepackage{graphicx}
\usepackage{booktabs} 

\usepackage[utf8]{inputenc} 
\usepackage[T1]{fontenc}    
\usepackage{url}            
\usepackage{booktabs}       
\usepackage{amsfonts}       
\usepackage{nicefrac}       
\usepackage{microtype}      
\usepackage{xcolor}         
\usepackage{enumitem}

\usepackage{amsmath}
\usepackage{amssymb}
\usepackage{graphicx}
\usepackage{multirow}
\usepackage{makecell}

\usepackage{caption}
\usepackage{subfig}
\usepackage{etoolbox}
\apptocmd{\thebibliography}{\setlength{\itemsep}{1pt}}{}{}

\newcommand{\model}{CleanUNet}

\title{Speech Denoising in the Waveform Domain with Self-Attention}
\name{Zhifeng Kong$^*$$^\ddag$~\thanks{$^*$Work done during an internship at NVIDIA.}, 
Wei Ping$^\dagger$, 
Ambrish Dantrey$^\dagger$,
Bryan Catanzaro$^\dagger$}
\address{ 
$^\ddag$UCSD \quad $^\dagger$NVIDIA \\
\tt{\hspace{-2cm}
z4kong@eng.ucsd.edu, 
  wping@nvidia.com}
}

\begin{document}

\maketitle

\begin{abstract}
In this work, we present \model, a causal speech denoising model on the raw waveform.
The proposed model is based on an encoder-decoder architecture combined with several self-attention blocks to refine its bottleneck representations, which is crucial to obtain good results.
The model is optimized through a set of losses defined over both waveform and multi-resolution spectrograms.
The proposed method outperforms the state-of-the-art models in terms of denoised speech quality from various objective and subjective evaluation metrics. ~\footnote{Audio samples: \url{https://cleanunet.github.io/}} We release our code and models at \url{https://github.com/nvidia/cleanunet}.
\end{abstract}

\section{Introduction}
\label{intro}
Speech denoising~\cite{loizou2007speech} is the task of removing background noise from recorded speech, while preserving the perceptual quality and intelligibility of speech signals.  It has important applications for audio calls, teleconferencing, speech recognition, and hearing aids, thus has been studied over several decades.
Traditional signal processing methods, such as spectral subtraction~\cite{boll1979suppression} and  Wiener filtering~\cite{lim1979enhancement}, are required to produce an estimate of noise spectra and then the clean speech given the additive noise assumption. 
These methods work well with stationary noise process, but may not generalize well to non-stationary or structured noise types such as dogs barking, baby crying, or traffic horn.

Neural networks have been introduced for speech enhancement since the 1980s~\cite{tamura1988noise, parveen2004speech}.
With the increased computation power, deep neural networks~(DNN) are often used, e.g.,~\cite{lu2013speech, xu2014regression}.
Over the past years, DNN-based methods for speech enhancement have obtained the state-of-the-art results.
The models are usually trained in the supervised setting, where the models learn to predict the clean speech signal given the input noise speech. 
These methods can be categorized into time-frequency and waveform domain methods.
A large body of works are based on time-frequency representations~(e.g., STFT)~\cite{xu2014regression,soni2018time, fu2019metricgan, fu2021metricgan+, wang2015deep, weninger2015speech, nicolson2019deep, germain2018speech, xu2017multi,
hao2021fullsubnet, westhausen2020dual, isik2020poconet}. They take the noisy spectral feature (e.g., magnitude of spectrogram, complex spectrum) as the input, and predict the spectral feature or certain mask for modulation of clean speech~(e.g., Ideal Ratio Mask~\cite{williamson2015complex}). Then, the waveform is resynthesized given the predicted spectral feature with other extracted information from the input noisy speech~(e.g., phase).
%
Another family of speech denoising models directly predict the clean speech waveform from the noisy waveform input~\cite{pascual2017segan, rethage2018wavenet, pandey2019tcnn, hao2019unetgan, defossez2020real}. 
In previous studies, although the waveform methods are conceptually compelling and sometimes preferred in subjective evaluation, they are still lagging behind the time-frequency methods in terms of objective evaluations~(e.g., PESQ~\cite{PESQ2001}). 

For waveform methods, there are two popular architecture backbones: WaveNet~\cite{oord2016wavenet} and U-Net~\cite{ronneberger2015u}. 
WaveNet can be applied to process the raw waveform~\cite{rethage2018wavenet} as in speech synthesis~\cite{ping2018clarinet}, or low-dimensional embeddings for efficiency reason~\cite{luo2019conv}.
U-Net is an encoder-decoder architecture with skip connections tailored for dense prediction, which is particularly suitable for speech denoising~\cite{pascual2017segan}.
To further improve the denoising quality, different bottleneck layers between the encoder and the decoder are introduced. Examples include dilated convolutions~\cite{stoller2018wave,hao2019unetgan}, stand-alone WaveNet~\cite{pandey2019tcnn}, and LSTM~\cite{defossez2020real}.
In addition, several works use the GAN loss to train their models \cite{pascual2017segan, baby2019sergan, hao2019unetgan, phan2020improving}.

In this paper, we propose \model, a causal speech denoising model in the waveform domain.
Our model is built on a U-Net structure~\cite{ronneberger2015u}, consisting of an encoder-decoder with skip connections and bottleneck layers. 
In particular,  we use masked self-attentions~\cite{vaswani2017attention} to refine the bottleneck representation, which is crucial to achieve the state-of-the-art results.
Both the encoder and decoder only use causal convolutions,
so the overall architectural latency of the model is solely determined
 by the temporal resampling ratio between the original
time-domain waveform and the bottleneck representation~(e.g., 16ms for 16kHz audio).
We conduct a series of ablation studies on the architecture and loss function, and demonstrate that our model outperforms the state-of-the-art waveform methods in terms of various evaluations.

\vspace{-.2em}
\section{Model}
\label{model}
\vspace{-.4em}
\subsection{Problem Settings}
\label{problem settings}
\vspace{-.4em}
The goal of this paper is to develop a model that can denoise speech collected from a single channel microphone. The model is causal for online streaming applications.
Formally, let $x_{\rm{noisy}} \in R^T$ be the observed noisy speech waveform of length $T$. We assume the noisy speech is a mixture of clean speech $x$ and background noise $x_{\rm{noise}}$ of the same length: $x_{\rm{noisy}} = x + x_{\rm{noise}}$. The goal is to obtain a denoiser function $f$ such that (1) $\hat{x} = f(x_{\rm{noisy}}) \approx x$, and (2) $f$ is causal: the $t$-th element of the output $\hat{x}_t$ is only a function of previous observations $x_{1:t}$.

\vspace{-.3em}
\subsection{Architecture}
\label{architecture}
\begin{figure}[!t]
    \centering
    \includegraphics[trim=205 120 265 150, clip, width=0.99\linewidth]{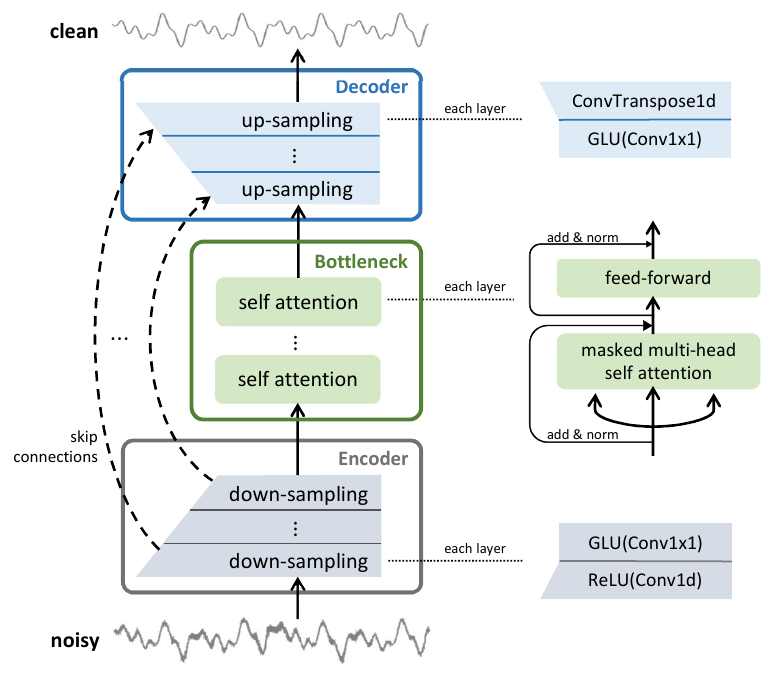}
    \vspace{-.4em}
    \caption{Model architecture for \model. It is causal with noisy waveform as input and clean speech waveform as output.}
    \label{fig:architecture}
\end{figure}

\vspace{-.3em}
We adopt a U-Net architecture \cite{ronneberger2015u, petit2021u} for our model $f$. It contains an encoder, a decoder, and a bottleneck between them. The encoder and decoder have the same number of layers, and are connected via skip connections (identity maps). We visualize the model architecture in Fig. \ref{fig:architecture}. 

\noindent\textbf{Encoder.} The encoder has $D$ encoder layers, where $D$ is the depth of the encoder. Each encoder layer is composed of a strided 1-d convolution (Conv1d) followed by the rectified linear unit (ReLU) and an 1$\times$1 convolution (Conv1$\times$1) followed by the gated linear unit (GLU), similar to \cite{defossez2020real}. The Conv1d down-samples the input in length and increases the number of channels. The Conv1d in the first layer outputs $H$ channels, where $H$ controls the capacity of the model, and Conv1d's in other layers double the number of channels. Each Conv1d has kernel size $K$ and stride $S=K/2$. 
Each Conv1$\times$1 doubles the number of channels because GLU halves the channels. All 1-d convolutions are causal.

\noindent\textbf{Decoder.} The decoder has $D$ decoder layers. Each decoder layer is paired with the corresponding encoder layer in the reversed order; for instance, the last decoder layer is paired with the first encoder layer. Each pair of encoder and decoder layers are connected via a skip connection. Each decoder layer is composed of a Conv1$\times$1 followed by GLU and a transposed 1-d convolution (ConvTranspose1d). The ConvTranspose1d in each decoder layer is causal and has the same hyper-parameters as the Conv1d in the paired encoder layer, except that the number of input and output channels are reversed. 

\noindent\textbf{Bottleneck.} The bottleneck is composed of $N$ self attention blocks \cite{vaswani2017attention}. Each self attention block is composed of a multi-head self attention layer and a position-wise fully connected layer. Each layer has a skip connection around it followed by layer normalization. Each multi-head self attention layer has 8 heads, model dimension 512, and the attention map is masked to ensure causality. Each fully connected layer has input channel size 512 and output channel size 2048.

\vspace{-.3em}
\subsection{Loss Function}
\label{loss}
\vspace{-.3em}
The loss function has two terms: the $\ell_1$ loss on waveform and an STFT loss between clean speech $x$ and denoised speech $\hat{x} = f(x_{\rm{noisy}})$. Let $s(x;\theta) = |\mathrm{STFT}(x)|$ be the magnitude of the linear-scale spectrogram of $x$, where $\theta$ represents the hyperparameters of STFT including the hop size, the window length, and the FFT bin. Then, we use multi-resolution STFT~\cite{yamamoto2020parallel} as our full-band STFT loss:
\begin{equation}
\label{eq: full band stft}
\begin{array}{l}
    \displaystyle \mathrm{M\text{-}STFT}(x,\hat{x}) = \\
    ~~~~~~ \sum_{i=1}^m
    \left( \frac{\|s(x;\theta_i)-s(\hat{x};\theta_i)\|_F}{\|s(x;\theta_i)\|_F} \right. + \left. \frac{1}{T} {\left\|\log \frac{s(x;\theta_i)}{s(\hat{x};\theta_i)}\right\|_1} \right),
\end{array}
\end{equation}
where $m$ is the number of resolutions and $\theta_i$ is the STFT parameter for each resolution. 
Then, our loss function is $\frac12\mathrm{M\text{-}STFT}(x,\hat{x}) + \|x-\hat{x}\|_1$.

In practice, we find the full-band $\mathrm{M\text{-}STFT}$ loss sometimes leads to low-frequency noises on the silence part of denoised speech, which deteriorates the human listening test.
On the other hand, if our model is trained with $\ell_1$ loss only, the silence part of the output is clean, but the high-frequency bands are less accurate than the model trained with the $\mathrm{M\text{-}STFT}$ loss.
This motivated us to define a high-band multi-resolution STFT loss.
Let ${s}_h(x)$ only contains the second half number of rows of $s(x)$~(e.g.,  4kHz to 8kHz range of the frequency bands for 16kHz audio). Then, the high-band STFT loss $\mathrm{M\text{-}{STFT}}_{h} (x,\hat{x})$ is defined by substituting $s(\cdot)$ with ${s}_h(\cdot)$ in Eq.~\eqref{eq: full band stft}.
Finally, the high-band loss function is $\frac12\mathrm{M\text{-}{STFT}}_h (x,\hat{x}) + \|x-\hat{x}\|_1$.

\begin{table*}[!t]\small
    \centering
    \caption{ Objective and subjective evaluation results for denoising on the DNS no-reverb testset.}
    \vspace{-.7em}
    \begin{tabular}{l|c|ccc|ccc|ccc}
    \toprule
        \multirow{2}{*}{Model} & \multirow{2}{*}{Domain} & PESQ & PESQ & STOI & pred. & pred. & pred. & MOS & MOS & MOS \\
        & & (WB) & (NB) & (\%) & CSIG & CBAK & COVRL & SIG & BAK & OVRL \\ \hline
        Noisy dataset & -           & 1.585 & 2.164 & 91.6 & 3.190 & 2.533 & 2.353 & -   & -   & - \\ \hline 
        DTLN \cite{westhausen2020dual}          & Time-Freq   & -     & 3.04  & 94.8 & -     & -     & -     & -   & -   & - \\
        PoCoNet \cite{isik2020poconet}       & Time-Freq   & 2.745 & -     & -    & 4.080 & 3.040 & 3.422 & -   & -   & - \\
        FullSubNet \cite{hao2021fullsubnet}    & Time-Freq   & 2.897 & 3.374 & 96.4 & 4.278 & 3.644 & 3.607 & 3.97  & 3.72 & 3.75\\
        Conv-TasNet \cite{luo2019conv}   & Waveform    & 2.73  & -     & -    & -     & -     & -     & -   & -   & - \\
        FAIR-denoiser~\cite{defossez2020real} & Waveform    & 2.659 & 3.229 & 96.6 & 4.145 & 3.627 & 3.419 & 3.68 & \bf{4.10} &  3.72\\ \hline
        \model~($N=5$, $\ell_1$+full) & Waveform    & \bf{3.146} & \bf{3.551} & \bf{97.7} & \bf{4.619} & \bf{3.892} & \bf{3.932} & \bf{4.03} & 3.89 & 3.78 \\
        \model~($N=5$, $\ell_1$+high) & Waveform    & 3.011 & 3.460 & 97.3 & 4.513 & 3.812 & 3.800 & 3.94 & \bf{4.08} &  \bf{3.87}\\
    \bottomrule
    \end{tabular}
    \label{tab: dns}
    \vspace{-.3em}
\end{table*}
\begin{table*}[!t]\small
    \centering
    \caption{Objective evaluation results for denoising on the Valentini test dataset.}
    \vspace{-.7em}
    \begin{tabular}{l|c|ccc|ccc}
    \toprule
        \multirow{2}{*}{Model} & \multirow{2}{*}{Domain} & PESQ & PESQ & STOI & pred. & pred. & pred.  \\
        & & (WB) & (NB) & (\%) & CSIG & CBAK & COVRL  \\ \hline
        Noisy dataset   & - &   1.943 & 2.553 & 92.7 & 3.185 & 2.475 & 2.536 \\ \hline
        FAIR-denoiser \cite{defossez2020real} & Waveform & 2.835 & 3.329 & 95.3 & 4.269 & 3.377 & 3.571 \\ \hline
        \model~($N=3$, $\ell_1$+full) & Waveform  & 2.884 & 3.396 & 95.5 & 4.317 & 3.406 & 3.625 \\
        \model~($N=5$, $\ell_1$+full) & Waveform   & \textbf{2.905} & \textbf{3.408} & \textbf{95.6} & \textbf{4.338} & \textbf{3.422} & \textbf{3.645} \\
    \bottomrule
    \end{tabular}
    \label{tab: valentini}
     \vspace{-.3em}
\end{table*}
\begin{table*}[!t]\small
    \centering
    \caption{ Objective evaluation results for denoising on the internal test dataset.}
    \vspace{-.7em}
    \begin{tabular}{l|c|ccc|ccc}
    \toprule
        \multirow{2}{*}{Model} & \multirow{2}{*}{Domain} & PESQ & PESQ & STOI & pred. & pred. & pred.  \\
        & & (WB) & (NB) & (\%) & CSIG & CBAK & COVRL  \\ \hline
        Noisy dataset           & -              & 1.488 & 1.861 & 86.2 & 2.436 & 2.398 & 1.927  \\ \hline 
        FullSubNet \cite{hao2021fullsubnet}        & Time-Freq      & 2.442 & 2.908 & 92.0 & 3.650 & 3.369 & 3.046  \\
        Conv-TasNet \cite{luo2019conv}             & Waveform       & 1.927 & 2.349 & 92.2 & 2.869 & 2.004 & 2.388  \\
        FAIR-denoiser \cite{defossez2020real}     & Waveform       & 2.087 & 2.490 & 92.8 & 3.120 & 3.173 & 2.593  \\ \hline
        \model~($N=5$, $\ell_1$+full) & Waveform       & \bf{2.450} & \bf{2.915} & \bf{93.8} & \bf{3.981} & \bf{3.423} & \bf{3.270}  \\
        \model~($N=5$, $\ell_1$+high) & Waveform       & 2.387 & 2.833 & 93.3 & 3.905 & 3.347 & 3.151  \\
    \bottomrule
    \end{tabular}
    \label{tab: rtx}
     \vspace{-.3em}
\end{table*}

\vspace{-.2em}
\section{Experiment}
\label{experiment}
\vspace{-.3em}
We evaluate the proposed \model~on three datasets: the Deep Noise Suppression (DNS) dataset \cite{reddy2020interspeech}, the Valentini dataset  \cite{ValentiniBotinhao2017NoisySD}, and an internal dataset. We compare our model with the state-of-the-art~(SOTA)  models in terms of various objective and subjective evaluations.
The results are summarized in Tables~\ref{tab: dns}, \ref{tab: valentini}, and \ref{tab: rtx}.
We also perform an ablation study for the architecture design and  loss function.

\vspace{.1em}
\noindent\textbf{Data preparation:} For each training data pair $(x_{\rm{noisy}}, x)$, we randomly take (aligned) $L$-second clips for both of them. We then apply various data augmentation methods to these clips, which we will specify for each dataset. 

\noindent\textbf{Architecture:} The depth of the encoder or the decoder is $D=8$. Each encoder/decoder layer has hidden dimension $H=48$ or $64$, stride $S=2$, and kernel size $K=4$. The depth of the bottleneck is $N=3$ or $5$. Each self attention block has 8 heads, model dimension $=512$, middle dimension $=2048$, no dropout and no positional encoding. 

\noindent\textbf{Optimization:} We use the Adam optimizer with momentum $\beta_1 = 0.9$ and denominator momentum $\beta_2 = 0.999$. We use the linear warmup with cosine annealing learning rate schedule with maximum learning rate $=2\times10^{-4}$ and warmup ratio $=5\%$. We survey three kinds of losses: (1) $\ell_1$ loss on waveform, (2) $\ell_1$ loss + full-band STFT, (3) $\ell_1$ loss + high-band STFT described in Section \ref{loss}. All models are trained on 8 NVIDIA V100 GPUs. 

\noindent\textbf{Evaluation:} We conduct both objective and subjective evaluations for denoised speech. Objective evaluation methods include (1) Perceptual Evaluation of Speech Quality (PESQ \cite{PESQ2001}), (2) Short-Time Objective Intelligibility (STOI \cite{taal2011algorithm}), and (3) Mean Opinion Score (MOS) prediction of the $i$) distortion of speech signal~(SIG), $ii$) intrusiveness of background noise (BAK), and $iii$) overall quality (OVRL) \cite{4389058}. 
For subjective evaluation, we perform a MOS test as
recommended in ITU-T P.835~\cite{recommendation2003subjective}.
We launched a crowd source evaluation on Mechanical Turk.
We randomly select 100 utterances from the test set, and each utterance is scored by 15 workers along three axis:   SIG, BAK, and OVRL.

\vspace{-.3em}
\subsection{DNS}
\label{dns}
\vspace{-.3em}
The DNS dataset \cite{reddy2020interspeech} contains recordings of more than 10K speakers reading 10K books with a sampling rate $16$kHz. We synthesize 500 hours of clean \& noisy speech pairs with 31 SNR levels (-5 to 25 dB) to create the training set \cite{reddy2020interspeech}. We set the clip $L=10$. We then apply the RevEcho augmentation \cite{defossez2020real} (which adds decaying echoes) to the training data. 

We compare the proposed \model~with several SOTA models on DNS.
By default, FAIR-denoiser~\cite{defossez2020real} only use  $\ell_1$ waveform loss on DNS dataset for better subjective evaluation.
We use hidden dimension $H=64$. We study both full and high-band STFT losses as described in Section \ref{loss} with multi-resolution hop sizes in $\{50,120,240\}$, window lengths in $\{240, 600, 1200\}$, and FFT bins in $\{512, 1024, 2048\}$. We train the models for 1M iterations with a batch size of $16$. We report the objective and subjective evaluations on the no-reverb testset in Table~\ref{tab: dns}. 
\model~ outperforms all baselines in objective evaluations.
For MOS evaluations, it get the highest SIG and OVRL, and slightly lower BAK.
\vspace{.2em}
\noindent\textbf{Ablation Studies:}
We perform ablation study to investigate the hyper-parameters and loss functions in our model. 
We survey $N=3$ or $5$ self attention blocks in the bottleneck.  We also survey different combination of loss terms. The objective evaluation results are in Table \ref{tab: dns ablation params}.  Full-band STFT loss and $N=5$ consistently outperforms the others in terms of objective evaluations.

\begin{table}[!t]\small
    \centering
    \caption{Objective evaluation results for the ablation study of \model~on the DNS no-reverb testset.}
    \vspace{-.7em}
    \begin{tabular}{cc|ccc}
    \toprule
        $N$ & loss  & PESQ (WB) & PESQ (NB) & STOI (\%) \\ \hline
        \multirow{3}{*}{3}
        & $\ell_1$           & 2.750 & 3.319 & 96.9 \\ 
        & $\ell_1$ + full    & 3.128 & 3.539 & 97.6 \\
        & $\ell_1$ + high    & 3.006 & 3.453 & 97.3 \\ \hline
        \multirow{3}{*}{5}
        & $\ell_1$     & 2.797 & 3.362 & 97.0 \\ 
        & $\ell_1$ + full    & 3.146 & 3.551 & 97.7 \\ 
        & $\ell_1$ + high    & 3.011 & 3.460 & 97.3 \\
    \bottomrule
    \end{tabular}
    \label{tab: dns ablation params}
\end{table}
\begin{table}[!t]\small
    \centering
    \caption{Objective evaluation results for the ablation study of architecture on the DNS no-reverb testset. Stride is $S=K/2$. RS means resampling using  sinc interpolation~\cite{smith1984flexible}. All models downsample 16khz audio 256$\times$ for bottleneck representation.}
     \vspace{-.7em}
    \begin{tabular}{lccc|ccc}
    \toprule
        \multirow{2}{*}{Model} & \multirow{2}{*}{$D$} & \multirow{2}{*}{$K$} & \multirow{2}{*}{RS} & PESQ & PESQ & STOI \\
        &&&& (WB) & (NB) & (\%) \\ \hline
        FAIR-denoiser & 5 & 8 & yes & 2.849 & 3.329 & 96.9 \\
        \model~& 5 & 8 & yes & 3.024 & 3.459  & 97.3 \\
        \model~& 4 & 8 & no  & 2.982 & 3.425 & 97.2 \\
        \model~& 8 & 4 & no  & 3.146 & 3.551 & 97.7 \\ 
    \bottomrule
    \end{tabular}
    \label{tab: dns ablation fair}
\end{table}

In order to show the importance of each component in \model, we perform an ablation study by gradually replacing the components in FAIR-denoiser~\cite{defossez2020real} to our model in Table \ref{tab: dns ablation fair}. First, we train FAIR-denoiser with the full-band STFT loss. The architecture has $D=5$ (kernel size $K=8$ and stride $S=4$), a resampling layer (realized by the sinc interpolation filter \cite{smith1984flexible}), and LSTM as the bottleneck.
Next, we replace the LSTM with $N=5$ self attention blocks, which leads to a boost of PESQ score from 2.849 to 3.024. 
We then remove the sinc resampling operation and reduce the depth of the encoder/decoder to $D=4$, which only has minor impact. 
Finally, we double the depth to $D=8$ and reduce the kernel size to $K=4$ and stride to $S=2$, which significantly improves the results. 
Note that all these models produce the same length  of bottleneck representation.

\vspace{.2em}
\noindent\textbf{Inference Speed and Model Footprints:}
We compare the inference speed and model footprints between the FAIR-denoiser and \model. We use hidden dimension $H=64$ in all models. The inference speed is measured in terms of the real time factor (RTF), which is the time required to generate certain speech divided by the total time of the speech. We use a batch-size of $4$ and length $L=10$ seconds under a sampling rate 16kHz. The results are reported in Table \ref{tab: stat}. 

\begin{table}[!t]\small
    \centering
    \caption{Inference speed (RTF) and model footprints (\#/ parameters) of the FAIR-denoiser and \model.
    The algorithmic latency for all models is 16ms for 16khz audio.}
      \vspace{-.7em}
    \begin{tabular}{l|cc}
    \toprule
        Model & RTF & Footprint \\ \hline
        FAIR-denoiser & $2.59\times10^{-3}$ & 33.53M \\
        \model~($N=3$) & $3.19\times10^{-3}$ & 39.77M \\
        \model~($N=5$) & $3.43\times10^{-3}$ & 46.07M \\
    \bottomrule
    \end{tabular}
    \label{tab: stat}
    \vspace{-.4em}
\end{table}

\vspace{-.3em}
\subsection{Valentini}
\label{valentini}
\vspace{-.4em}
The Valentini dataset \cite{ValentiniBotinhao2017NoisySD} contains 28.4 hours of clean \& noisy speech pairs collected from 84 speakers with a sampling rate $48$kHz. There are 4 SNR levels (0, 5, 10, and 15 dB). At each iteration, we randomly select $L$ to be a real value between 1.33 and 1.5. We then apply the Remix augmentation  (which shuffles noises within a local batch) and the BandMask augmentation \cite{defossez2020real} (which removes a fraction of frequencies) to the training data.
We compare our \model~ with FAIR-denoiser \cite{defossez2020real} as the baseline. In both models we use hidden dimension $H=48$ and the full-band STFT loss with the same setup as DNS dataset. We train the models for 1M iterations with a batch size of $64$. We report the objective evaluation results in Table \ref{tab: valentini}. 
Our model consistently outperforms FAIR-denoiser in all metrics.

\vspace{-.3em}
\subsection{Internal Dataset}
\label{rtx}
\vspace{-.4em}
We also conduct experiments on an challenging internal dataset. The dataset contains 100 hours of clean \& noisy speech with a sampling rate $48$kHz. Each clean audio may contain speech from multiple speakers. We use the same data preparation ($L=10$) and augmentation method as the DNS dataset in Section \ref{dns}. 

We compare the proposed \model~with FAIR-denoiser and FullSubNet on this internal dataset. We set $H=64$, $N=5$, $D=8$, $K=8$, $S=2$. We use full and high-band STFT losses with the same setup as DNS dataset. We train the models for 1M iterations with a batch size of $16$. We report the objective evaluation results in Table \ref{tab: rtx}. 
Our model consistently outperforms FullSubNet and FAIR-denoiser in all metrics.

\section{Conclusion}

In this paper, we introduce \model, a causal speech denoising model in the waveform domain. 
Our model uses the U-Net like encoder-decoder as the backbone, and relies on self-attention modules to refine its bottleneck representation.
We test our model on various datasets, and show that it achieves the state-of-the-art performance in terms of objective and subjective evaluation metrics in the speech denoising task. We also conduct a series of ablation studies for architecture design choices and different combinations of loss functions.

\bibliographystyle{IEEEbib}
{\small
\bibliography{main}
}
\end{document}